\documentclass[a4paper,11pt]{article}
\usepackage{graphicx}
\usepackage{epsf,amsmath,bbold,amsfonts,stmaryrd}

\usepackage[utf8]{inputenc}
\usepackage{mathrsfs}
\usepackage{appendix}
\usepackage{amssymb}
\usepackage{float}
\usepackage{color}
\usepackage{cite}
\usepackage{hyperref}
\hypersetup{pageanchor=false}
\usepackage{indentfirst}
\usepackage{url}
\usepackage{float}
\usepackage{caption}
\usepackage[numbers,square,comma,sort&compress,merge]{natbib}

\def\ep{\epsilon}

\def\mc{\mathcal}

\def\nn{\nonumber}

\def\t{\theta}

\def\be{\begin{equation}}
\def\ee{\end{equation}}

\def\bea{\begin{eqnarray}}
\def\eea{\end{eqnarray}}

\def\ba{\begin{array}}
\def\ea{\end{array}}

\def\bc{\begin{center}}
\def\ec{\end{center}}

\def\bl{\begin{flushleft}}
\def\el{\end{flushleft}}

\def\br{\begin{flushright}}
\def\er{\end{flushright}}

\def\bi{\begin{itemize}}
\def\ei{\end{itemize}}

\def\bt{\begin{tabular}}
\def\et{\end{tabular}}

\numberwithin{equation}{section}

\numberwithin{equation}{section}

\begin{document}
\title{\textbf{High spin expansion for null geodesics}}
\author{Peng-Cheng Li$^{1,2}$,
Minyong Guo$^{1*}$,
and
Bin Chen$^{1,2,3}$}
\date{}

\maketitle

\vspace{-10mm}

\begin{center}
{\it
$^1$Center for High Energy Physics, Peking University,
No.5 Yiheyuan Rd, Beijing 100871, P. R. China\\\vspace{1mm}

$^2$Department of Physics and State Key Laboratory of Nuclear
Physics and Technology, Peking University, No.5 Yiheyuan Rd, Beijing
100871, P.R. China\\\vspace{1mm}

$^3$ Collaborative Innovation Center of Quantum Matter,
No.5 Yiheyuan Rd, Beijing 100871, P. R. China
}
\end{center}

\vspace{8mm}

\begin{abstract}
{We consider the high spin expansion for the null geodesics in the Kerr spacetime. We expand the null geodesic equation successively to higher orders in deviation from extremality.  Via the method of matched asymptotic expansion, the radial integrals are obtained analytically. It turns out that the analytic expressions are very sensitive to the value of the shifted Carter constant $q$. We show that for a large $q$, the analytic expressions can be used to study observational electromagnetic signatures for astrophysical black holes like M87*. However, for a small $q$, the high spin expansion method can only be applied to (near-) extreme black holes.}
\end{abstract}

\vfill{\footnotesize Email: lipch2019@pku.edu.cn,\,\,minyongguo@pku.edu.cn,\,bchen01@pku.edu.cn\\$~~~~~~*$ Corresponding author.}

\maketitle

\newpage
\section{Introduction}
The Event Horizon Telescope (EHT) had enough participants to enable the array to reach sufficient angular resolution to image the supermassive black hole in Sagittarius A* and M87* for the first time in April 2017 \cite{EHT}. From then on, a new page has been turned in astronomical black hole physics. The large amount of available experimental data brings not only excitement but also challenges. In other words, we need not only to process data, but also to make theoretical predictions about what the EHT will see. In April 2019, the mystery of the image of M87* was firstly solved, which helps to test Einstein's general relativity and other gravity theories \cite{Akiyama:2019cqa, Akiyama:2019brx,Akiyama:2019sww,Akiyama:2019bqs,Akiyama:2019fyp,Akiyama:2019eap}. However, this is only the first step and the next is to increase the number of telescopes to take pictures of the black holes in more detail and from more angles. Correspondingly, further theoretical researches  have to be  developed.

As we know, the basic theoretical aspect of dealing with the observational electromagnetic signatures is the study of null geodesics in the black hole spacetime. In general, the integrals of the null geodesic equations are hard to perform analytically. Nevertheless, for a (near-) extreme Kerr black hole, the authors in \cite{Porfyriadis:2016gwb} found that the radial integrals
of the equations for null geodesics which start in the near-horizon region to an observer in the region far from the black hole can be solved analytically  to the leading order in $\ep$, a parameter which characterizes the deviation from extremality (see the definition eq.(\ref{defineepsilon})). The rationale behind the strategy is due to the emergence of the enhanced conformal symmetry in the near-horizon geometry for the (near-) extreme Kerr black hole \cite{Bardeen:1999px,Guica:2008mu,Guica:2008mu}.\footnote{The radial integrals of null geodesics can also be obtained analytically by using the $1/D$ expansion method, no matter the black holes are rotational or not \cite{Guo:2019pte}.} Recently, this enhanced symmetry helped to  simplify the gravitational dynamics greatly in the near-horizon region, see for instance \cite{Lupsasca:2017exc,Wei:2016avv, Guo:2018kis, Gates:2018hub, Long:2018tij, Yan:2019etp,Igata:2019pgb,Kapec:2019hro,Igata:2019hkz,Guo:201911,Compere:2020eat}.

As a prototypical example, the authors in \cite{Gralla:2017ufe} analytically computed the observational appearance of an emitter (viz., a point light source)
on the inner most stable orbit of a (near-) extreme Kerr black hole. Moreover, as pointed out in \cite{Gralla:2017ufe}, their results are valid only when the deviation parameter $\epsilon\leq0.01$, which corresponds to the spin $a\geq0.9999995 M$. However, to our knowledge, the spin of rapidly spinning black holes in our universe cannot be so large. Theoretically, the black holes with geometrically thin disk models are subjected to the well-known Thorne bound $a\leq 0.998 M$ \cite{Thorne:1974ve}. Concerning the observation of EHT, the target high spin black hole is M87*. Recent researches  have shown that the upper bound of the spin of M87* is about $0.9M$ \cite{Tamburini:2019vrf,Bambi:2019tjh}. In consequence, the approximate analytic results at the leading order of $\epsilon$ obtained by treating M87* as a (near-) extreme Kerr black hole may not be applicable anymore. A possible remedy to overcome this shortcoming  is to  perform the radial integrals of the null geodesic equations by taking into account the corrections of higher orders in $\ep$, such that the spin could take a value relevant to the realistic  supermassive black holes like M87*.

The aim of this paper is to investigate the feasibility of obtaining analytic results via expanding the dynamic equations of a rapidly spinning black hole in $\ep$ to high orders, which is called the high spin expansion. Based on the work of \cite{Porfyriadis:2016gwb,Gralla:2017ufe}, we would like to perform the high spin expansion for the radial integrals of the null geodesics in the Kerr spacetime. We first expand the null geodesic equations in $\ep$ successively including the leading order (LO), the next-to-leading order (NLO), and even up to NNNLO corrections.  With this, via the method of matched asymptotic expansion (MAE) we find that the radial integrals of the null geodesic equations can be obtained analytically order by order. The reliability of the approximate analytical results is crucially dependent on the values of the shifted Carter constant $q$. When $q$ is large, the results matches well with the numerical integrals for a relatively smaller spin, e.g. $a=0.9M$, which confirms the effectiveness of the high spin expansion. However, if $q$ takes a small value the high spin expansion method would be not so effective and the relative error of the approximate results to the numerical integrals increases quickly as the spin deviates from extremality. The explicit demonstration of the applications of the high spin expansion method in this paper may provide some hints in solving other dynamical equations.

The rest of the paper is organized as follows. In Sec. \ref{section2}, we discuss the basis of null geodesics and the high spin expansion in the Kerr spacetime. In Sec. \ref{section3}, we focus on the radial integrals of the null geodesic equations in the $r-\theta$ motion. Next, we turn to the radial integrals in $r-\phi$ and $r-t$ motions in Sec. \ref{section4} and \ref{section5}, respectively. We provide concluding remarks in Sec. \ref{summary}.

\section{Null geodesics and the high spin expansion in Kerr spacetime}\label{section2}
In this section, we would like to start with a brief review of null geodesics in Kerr spacetime. Then we turn to discuss the basic idea behind exploring the higher corrections to the radial integrals along the null geodesics, in the virtue of the high spin expansion and matched asymptotic expansion (MAE) at some length. Actually, this program was first explored in \cite{Porfyriadis:2016gwb,Gralla:2017ufe}, where they stopped at the calculations at the leading order however. As a consequence, our investigation to the higher orders  can be seen as an extension along this line.

\subsection{Null geodesics in Kerr spacetime}

In terms of the Boyer-Lindquist coordinates we have the Kerr metric in this form
\begin{equation}\label{Kerrmetric}
  d s^2 = - \frac{\Delta}{\hat{\rho}^2} (d \hat{t} - a \sin^2 \theta d
  \hat{\phi})^2 + \frac{\sin^2 \theta}{\hat{\rho}^2} ((\hat{r}^2 + a^2) d
  \hat{\phi} - a d \hat{t})^2 + \frac{\hat{\rho}^2}{\Delta} d \hat{r}^2 +
  \hat{\rho}^2 d \theta^2,
\end{equation}
where
\begin{equation}
  \Delta = \hat{r}^2 - 2 M \hat{r} + a^2, \quad \hat{\rho}^2 = \hat{r}^2 + a^2
  \cos^2 \theta .
\end{equation}
This describes neutral rotating black holes of mass $M$ and angular momentum
$J = a M$.
The 4-momentum of a massless particle living in the spacetime takes the general form
\be
p^\mu=(\dot{\hat{t}},\dot{\hat{r}}, \dot{\theta}, \dot{\hat{\phi}}),
\ee
where the dot denotes the derivative with respect to the affine parameter. In general, the motion of a particle is governed by the Hamilton–Jacobi equation, and the null geodesic equations are completely integrable in Kerr spacetime because of four conserved quantities along the trajectory of each photon: the invariant mass $p^2 = 0$, the total energy $\hat{E} = - p_t$, the component of angular momentum parallel to the axis of symmetry $\hat{L} = p_{\phi}$, and the Carter constant $\hat{Q} = p_{\theta}^2 - \cos^2 \theta (a^2 p_t^2 - p_{\phi}^2 \csc^2 \theta)$.

The existence of the four conserved quantities enables that the integration of the null geodesic equations can be cast into the form \cite{Carter:1968rr}
\begin{eqnarray}
  \int^{\hat{r}} \frac{d \hat{r}'}{\sqrt{\hat{R}}} & = & \int^{\theta} \frac{d
  \theta'}{\sqrt{\hat{\Theta}}}, \\
  \hat{\phi} & = & \int^{\hat{r}} \frac{a \hat{E}  \hat{r}'^2 + (\hat{L} -
  a \hat{E}) (\Delta - a^2)}{\hat{\Delta} \sqrt{\hat{R}}} d \hat{r}' +
  \int^{\theta} \frac{\hat{L} \cot^2 \theta'}{\sqrt{\hat{\Theta}}} d \theta',
  \\
  \hat{t} & = & \int^{\hat{r}} \frac{ \hat{E}  \hat{r}'^2 (\hat{r}'^2
  + a^2) + a (\hat{L} - a \hat{E}) (\Delta - \hat{r}'^2 -
  a^2)}{\hat{\Delta} \sqrt{\hat{R}}} d \hat{r}' + \int^{\theta} \frac{a^2
  \hat{E} \cos^2 \theta'}{\sqrt{\hat{\Theta}}} d \theta',
\end{eqnarray}
where
\begin{eqnarray}
  \hat{R} & = & ( \hat{E}  (\hat{r}'^2 + a^2) - \hat{L} a)^2 - \Delta(r') (\hat{Q} +
  (\hat{L} - a \hat{E})^2), \\
  \hat{\Theta} & = & \hat{Q} - \cos^2 \theta' \left( \frac{\hat{L}^2}{\sin^2
  \theta'} - a^2 \hat{E}^2 \right) .
\end{eqnarray}
The trajectory of the photon is independent of its energy, thus for convenience we introduce two rescaled quantities,
\begin{equation}
\hat{\lambda} = \frac{\hat{L}}{\hat{E} }, \quad \hat{q} =
  \frac{\sqrt{\hat{Q}}}{\hat{E} } .
\end{equation}
Note that since $\hat{Q}=p_\t^2$ when $\t=\pi/2$, any photon passing through the equatorial plane must have a nonnegative Carter constant, and hence a real $\hat{q}$. To avoid the case that when $\hat{Q}$ is negative then  $\hat{q}$ is imaginary, our strategy is to constrain the emitter in  the equatorial plane.

\subsection{The high spin expansion for Kerr spacetime}

In this subsection, we move to the high spin expansion for the Kerr spacetime and discuss the method of MAE. Let us start to introduce a small parameter to represent the deviation of the black hole from being extreme,
\begin{equation}\label{defineepsilon}
  \ep^3 = 1 - \frac{a^2}{M^2}.
\end{equation}
In \cite{Gralla:2017ufe}, it was shown that for an emitter orbiting near a high spin black hole, such as on the innermost stable orbit (ISCO), $\hat{\lambda}$ is of the form
\be\label{sbound}
 \hat{\lambda}=2M+\mc O(\ep),
 \ee
 which is near the superradiant bound \footnote{The superradiant bound is given by $\hat{E}=\Omega_H \hat{L}$, so for a high spin black hole one finds $\hat{\lambda}=2M(1+\ep^{3/2}+\mc O(\ep^3)$).}.
 To characterize this feature, as in \cite{Porfyriadis:2016gwb} we introduce
 \be
 \hat{\lambda}=2M(1-\ep \lambda).
 \ee
Moreover, following \cite{Gralla:2017ufe} we introduce the shifted Carter constant
\be
q^2=3-\frac{\hat{q}^2}{M^2}.
\ee
It turns out that a positive $q^2$ guarantees a geodesic originating in the near region of the horizon can reach out to the
far asymptotically flat region. The non-negativity of $\Theta$ restricts the constant $q$ and $\lambda$ by the inequality \cite{Chandrasekhar:1983}
\be
-q^2+\left(\sqrt{1-\epsilon ^3}+2 \lambda  \epsilon -2\right)^2+3\geq0,
\ee
which in the high spin expansion is given by
\be
4-q^2-4\lambda\ep+4\lambda^2\ep^2+\ep^3+\mc O(\ep^4)\geq0.
\ee
It is convenient to introduce the dimensionless Bardeen-Horowitz coordinates
\begin{equation}\label{BHcoor}
  t = \frac{\hat{t}}{2 M}, \quad \phi = \hat{\phi} - \frac{\hat{t}}{2 M},
  \quad r = \frac{\hat{r} - M}{M} .
\end{equation}
In terms of these coordinates, up to the leading order in $\epsilon$, there is
\be\label{NHEK}
r=\ep  \bar{R},\qquad  \bar{R}=\mc O(1).
\ee
It was shown \cite{Bardeen:1999px,Guica:2008mu} that the near-horizon region of the extreme Kerr black hole (referred to as the NHEK geometry)
 admits an enhanced Killing symmetry $SL(2,\mathbb{R})\times U(1)$. This enhanced symmetry renders the radial integral of the geodesic equations in the extreme Kerr spacetime is analytically solvable \cite{Porfyriadis:2016gwb}. Moreover, Ref.\cite{Bredberg:2009pv} found that the same symmetry appears for the near-horizon region of the near-extreme Kerr black hole (referred to as near-NHEK), since the NHEK and near-NHEK geometries are diffeomorphic to each other. The so-called near-NHEK region also refers to the region of an extreme Kerr black hole closer to the event horizon than the NHEK region, that is
  \be
  r=\ep^{3/2} \bar{R},\qquad  \bar{R}=\mc O(1).
  \ee
  Therefore, the null geodesic equations in the near-extreme Kerr spacetime are analytically solvable as well \cite{Gralla:2017ufe}. When the metric is expanded up to higher orders in $\ep$, the enhanced symmetry may not exist anymore, especially for a Kerr black hole that has a finite deviation from the extremality. Nevertheless, if we treat the higher order terms of $\ep$ as small corrections of the near-extreme black hole, the null geodesic equations are expected to be analytically solvable. In the following we will demonstrate this point through detailed computations. Particularly, in certain cases we find that $\ep$ can even take a finite value while maintaining convergence of the integrals.

In terms of the coordinates (\ref{BHcoor}), the null geodesic equations become
\begin{eqnarray}
  \int^{r_f}_{r_n} \frac{d r'}{\sqrt{R}} & = & \int^{\theta_f}_{\theta_n}
  \frac{d \theta}{\sqrt{\Theta}},\\
  \phi_f - \phi_n & = & - \frac{1}{2}\int_{r_n}^{r_f}\frac{\Phi}{(r^2-\ep^3)\sqrt{R}}dr
  +\frac12\int_{\theta_n}^{\theta_f}\frac{\left(\epsilon ^3-1\right) \cos ^2\theta -4(\lambda  \epsilon -1) \cot ^2\theta }{\sqrt{\Theta}}d\t,\\
  t_f - t_n & = & \frac{1}{2}\int_{r_n}^{r_f}\frac{T}{(r^2-\ep^3)\sqrt{R}}dr
  +\frac12\int_{\theta_n}^{\theta_f}\frac{(1-\ep^3)\cos^2\t}{\sqrt{\Theta}}d\theta,
\end{eqnarray}
where
\bea
 R &=&\left((r+1)^2+2 \sqrt{1-\epsilon ^3} (\lambda  \epsilon -1)-\epsilon ^3+1\right)^2-\left(r^2-\epsilon ^3\right) \left(-q^2+\left(\sqrt{1-\epsilon ^3}+2 \lambda  \epsilon -2\right)^2+3\right),\nn\\
\Theta&=&3-q^2+(1-\epsilon ^3)\cos ^2\theta -4(\lambda  \epsilon -1)^2\cot ^2\theta,\nn\\
\Phi &=&(r+1) \left(r^3+3 r^2-r \left(\epsilon ^3-4 \lambda  \epsilon \right)+4 \lambda  \epsilon  \left(\sqrt{1-\epsilon ^3}-1\right)-3 \epsilon ^3-8 \sqrt{1-\epsilon ^3}+8\right),\nn\\
T&=&(r+1) \left(r^3+3 r^2-r \left(\epsilon ^3-4\right)+4 \lambda  \epsilon  \sqrt{1-\epsilon ^3}-3 \epsilon ^3-4 \sqrt{1-\epsilon ^3}+4\right).
\eea
The above integrals are of elliptic type \cite{Gralla:2019ceu} (see also \cite{Rauch:1994,Vazquez:2003zm,Dexter:2009}) and could be treated numerically. In this paper we will show that the integrals can be expanded as power-series of  small $\epsilon$ and at each order of $\ep$ the integrals can be performed analytically. But  here is a subtlety. Let us consider the radial integrals initiating from the near-horizon region,
and terminating at the far region,
which is
\be
r_f\gg \ep.
\ee
The expansion in $\ep$ works well for $r$ within the far region, but  may not be valid anymore for $r\ll1$, because that in the near-horizon region, $r$ could be as small as $\sim\ep$ or even smaller, which leads to the invalidness of the expansion in $\ep$. Following \cite{Porfyriadis:2016gwb,Gralla:2017ufe}, here we consider the photons are emitted from the NHEK region, then to guarantee that the geodesics can get all the way to the far region, the superradiant bound (\ref{sbound}) has to be respected, i.e. $\lambda\geq0$, otherwise the photons would run into the turning point outside the event horizon.\footnote{Photons could also be emitted from the near-NHEK region, however, since the ISCO lies in the NHEK region, no stable circular obit for a timelike particle exists in the near-NHEK region.} To evade this obstacle,  one can introduce an intermediate scale $\ep^p$ with $0<p<1$, satisfying
\be\label{verlapregion}
\ep\ll \ep^p\ll1,
\ee
to separate the integration regions.
For the integral in the first region $r_n\leq r\leq \ep^p$, with $r_n$ lying within the NHEK region, we  make the change of variable
\be
x=r/\ep,
\ee
such that in both regions the radial integrals can be appropriately expanded in $\ep$.

The intermediate scale bears double features, one is that $\ep^p\ll1$ so it is in the near-horizon region, but at the same time since $\ep^p\gg\ep$, it also lies in the far region. Therefore, via the method of matched asymptotic expansion  (MAE) one can perform the radial integrals of the null geodesic equations in the high spin expansion. More specifically, one first performs the
radial integral in the near-horizon region, $r_n\leq r\leq \ep^p$,
and in the far region, $\ep^p\leq r\leq r_f$, respectively. The matching of the solutions in the overlap regions $r\sim \ep^p$ eliminate the dependence on $p$.

\section{The radial integral for the $r-\theta$ motion}\label{section3}

Based on the discussion in the previous section, next we compute the radial integrals practically using the high spin expansion and the method of MAE. In this section, we focus on the radial integral
\be
I^\theta=\int^{r_f}_{r_n}\frac{d r}{\sqrt{R(r)}},\label{raidalinteral_rtheta}
\ee
for the $r-\theta$ motion, which can be separated into
\be
I^\theta=I^\theta_n+I^\theta_f,
\ee
with
\be
I^\theta_n=\int^{\ep^p}_{r_n} \frac{d r}{\sqrt{R(r)}},\quad
I^\theta_f=\int^{r_f}_{\ep^p} \frac{d r}{\sqrt{ R(r)}}.
\ee
Then both $I^\theta_n$ and $I^\theta_f$ can be expanded in a series of $\ep$, that is,
\be
I^\theta_{n,f}=I^{\theta(0)}_{n,f}+I^{\theta(1)}_{n,f}\ep+I^{\theta(2)}_{n,f}\ep^2+\dots.
\ee
For brevity, in the following we will use the notations for integral at each order of $\ep$, that is,
\be
I_n^{\theta(i)}=\mc F^{\theta(i)}_n(\ep^{p-1})-\mc F^{\theta(i)}_n(x_n),
\ee
and
\be
I_f^{\theta (i)}=\mc F^{\theta (i)}_f(r_f)-\mc F^{\theta (i)}_f(\ep^p),
\ee
where $i=0,1,2,3,\dots$ and $x_n=r_n/\ep$.

First of all, to the leading order of $\ep$, one can easily find
\be
I_n^{\theta(0)}=\int^{\ep^{p-1}}_{x_n}\frac{d x}{\sqrt{\mc R_n(x)}}=\mc F^{\theta (0)}_n(\ep^{p-1})-\mc F^{\theta (0)}_n(x_n),
\ee
\be
I^{\theta (0)}_f=\int^{r_f}_{\ep^p}\frac{d r}{\sqrt{\mc R_f(r)}}
=\mc F^{\theta (0)}_f(r_f)-\mc F^{\theta (0)}_f(\ep^p)
,
\ee
with
\be
\mc F^{\theta (0)}_n(x)=\frac{1}{q}\log \left(q \sqrt{\mc R_n(x)}+q^2 x+4 \lambda \right),
\ee
and
\be
\mc F^{\theta (0)}_f(r)=-\frac{1}{q}\log\frac{q\sqrt{\mc R_f(r)}+q^2 r+2r^2}{r^2},
\ee
where we have introduced
\be
\mc R_n(x)=q^2 x^2+4 \lambda  (\lambda +2 x),
\ee
\be
\mc R_f(r)=r^2 (r^2+4r+q^2).
\ee
One can see that both the upper limit of the radial integral $I^{(0)}_n$, i.e. $\mc F^{\theta (0)}_n(\ep^{p-1})$, and the lower limit of $I^{\theta (0)}_f$, i.e. $\mc F^{\theta (0)}_f(\ep^p)$,
depend on the specific value of $p$. But one can find that the asymptotic
form  of $\mc F^{\theta (0)}_n(\ep^{p-1})-\mc F^{\theta (0)}_f(\ep^p)$ is independent of $p$, which is essentially the application of the MAE.
In detail, since $\ep^{p-1}\gg1$, one can expand $\mc F^{\theta (0)}_n(\ep^{p-1})$ as
\be
\mc F^{\theta (0)}_n(\ep^{p-1})=\frac{\log (2q^2)-(1-p)\log \ep}{q}+\frac{4 \lambda }{q^3}\ep^{1-p}+\frac{ \lambda ^2 \left(q^2-12\right)}{q^5}\ep^{2(1-p)}
+\mc O(\ep^{3(1-p)}),
\ee
and since $\ep^p\ll1$, one can expand $\mc F^{\theta (0)}_f(\ep^p)$ as
\be
\mc F^{\theta (0)}_f(\ep^p)=-\frac{\log (2q^2)-p\log \ep}{q}-\frac{2 }{q^3}\ep^p-\frac{\left(q^2-12\right) }{4 q^5}\ep^{2p}+\mc O(\ep^{3p}).
\ee
We can see that the leading order of $\mc F^{\theta (0)}_n(\ep^{p-1})-\mc F^{\theta (0)}_f(\ep^p)$ is indeed independent of $p$, with the higher order terms can be canceled by the leading order terms of the higher order $\mc F$ functions. For example, the second term in $\mc F^{\theta (0)}_n(\ep^{p-1})$ can counteract the first term in
$\mc F^{\theta (1)}_f(\ep^p)$.
Then as shown in \cite{Gralla:2017ufe}, one obtains
\be\label{rtheta0}
I^{\theta (0)}=-\frac{1}{q}\log\ep+\frac{1}{q}\log\frac{4q^4r_f^2}
{(q \sqrt{\mc R_n(x_n)}+q^2 x_n +4 \lambda )(q\sqrt{\mc R_f(r_f)}+q^2 r_f+2r_f^2)}.
\ee
The above procedure works straightforwardly when we proceed to the radial integrals
expanded at higher orders in $\ep$, i.e. $I^{\theta (i)}$  for $i\geq2$.
In order to obtain the form independent of $p$ for the integral $I^{i}$, one has to expand the upper limits of all $I^{\theta (1)}_n$, $I^{(2)}_n$, $\dots$, $I^{\theta (i)}_n$ around $\ep^{p-1}$ to enough orders and  the same thing occurs for the lower limits of all $I^{\theta (1)}_f$, $I^{\theta (2)}_f$, $\dots$, $I^{\theta (i)}_f$ around $\ep^{p}$. Now let us do this step by step.

The radial integrals at the next-to-leading order are analytically obtained as
\be
I^{\theta (1)}_n=\int^{\ep^{p-1}}_{x_n}\frac{q^2-4(1+2 \lambda x^2 +x^3)}{2 \mc R^{3/2}_n}dx=\mc F^{\theta (1)}_n(\ep^{p-1})-\mc F^{\theta (1)}_n(x_n),
\ee
with
\bea
\mc F^{\theta (1)}_n(x)&=&\frac{1}{8 \lambda ^2 q^5 \left(q^2-4\right) \sqrt{\mc R_n}}\Bigg[
q^9 x-4 q^7 (x-\lambda )+16 \lambda  q^5 \left(-\lambda  x^2+2 \lambda ^2 x-1\right)+64 \lambda ^2 q^3 \left(-4 \lambda ^2+x^2-9 \lambda  x\right)\nonumber\\
&&-32 \lambda ^3 \left(q^4-10 q^2+24\right) \sqrt{\mc R_n} \log \left(4 \lambda +q^2 x+q \sqrt{\mc R_n}\right)+768 \lambda ^3 q (\lambda +2 x)\Biggl],
\eea
and
\be
I^{\theta (1)}_f=\int^{r_f}_{\ep^p}-\frac{4 (\lambda  r (r+1))}{\mc R_f^{3/2}}dr
=\mc F^{\theta (1)}_f(r_f)-\mc F^{\theta (1)}_f(\ep^p),
\ee
with
\bea
\mc F^{\theta (1)}_f(r)&=&\frac{4\lambda}{q^5 \left(q^2-4\right) r \sqrt{\mc R_f}}\Bigg[-\left(q^4-10 q^2+24\right) r \sqrt{\mc R_f} \log \left(\frac{q \sqrt{\mc R_f}+q^2r+2 r^2}{r^2}\right)+12 q r (r+4)\nonumber\\
&&+q^5 (r-1)-2 q^3 \left(2 r^2+9 r-2\right)\Bigg].
\eea
As before, both $\mc F^{\theta (1)}_n(\ep^{p-1})$ and $\mc F^{\theta (i)}_f(\ep^p)$ depend on the specific value of $p$, but via the MAE, those $p$-dependent terms are properly canceled out by the higher order $\mc F$ functions. Then one can extract the $p$-independent terms
\be
I^{\theta (1)}=\mc F^{\theta (1)}_f(r_f)-\mc F^{\theta (1)}_n(x_n)+\frac{4 \lambda  \left(q^2-6\right) }{q^5}\log \epsilon
-\frac{8 \lambda  \left(q^2-6\right) \log \left(2 q^2\right)}{q^5}+\frac{8 \lambda  \left(q^4-16 q^2+40\right)}{q^5 \left(q^2-4\right)}+\frac{q}{8 \lambda ^2}.
\ee
Now, let us move on to the next order of the $\ep$ expansion. The radial integrals at the next-to-next-to-leading order (NNLO) are analytically obtained as
\bea
I^{\theta (2)}_n&=&\int^{\ep^{p-1}}_{x_n}\frac{3 \left(q^2-4 \left(x^3+2 \lambda  x^2+1\right)\right)^2-4\mc R_n \left(-4 \lambda +x^4-4 \lambda ^2 x^2\right)}{8 \mc R_n^{5/2}} dx\nonumber\\
&=&\mc F^{\theta (2)}_n(\ep^{p-1})-\mc F^{\theta (2)}_n(x_n),
\eea
 and
 \bea
 I^{\theta (2)}_f&=&\int^{r_f}_{\ep^p}\frac{2 \lambda ^2 (r+1) r^2 \left(q^2 (r-1)+r^3+3 r^2+8 r+12\right)}{\mc R_f^{5/2}} dr\nonumber\\
 &&=\mc F^{\theta (2)}_f(r_f)-\mc F^{\theta (2)}_f(\ep^p),
 \eea
 with
 \bea
 \mc F^{\theta (2)}_n(x)&=&\frac{1}{64 \lambda ^4 q^8 \left(q^2-4\right)^2 \mc R_n^{3/2}}\Bigg[q^{18} x^3+2 q^{16} x \left(3 \lambda ^2-4 x^2+6 \lambda  x\right)+8 q^{14} \left(3 \lambda ^3+2 x^3-12 \lambda  x^2-3 \lambda ^2 x\right)\nonumber\\
&& -16 \lambda  q^{12} \left(14 \lambda ^2+\lambda ^3 x^5+\left(8 \lambda ^5-4 \lambda ^2\right) x^3-12 \left(\lambda ^3+1\right) x^2+\left(6 \lambda -8 \lambda ^4\right) x\right)\nonumber\\
&&-64 \lambda ^2 q^{10} \left(-10 \lambda -5 \lambda ^2 x^5-27 \lambda ^3 x^4+4 \lambda  \left(3 \lambda ^3+2\right) x^3+8 \lambda ^2 \left(\lambda ^3+3\right) x^2+2 \left(4 \lambda ^6+8 \lambda ^3-3\right) x\right)\nonumber\\
&&+256 \lambda ^3 q^8 \left(8 \lambda ^6-7 \lambda  x^5-75 \lambda ^2 x^4+\left(418 \lambda ^3+4\right) x^3+4 \left(37 \lambda ^3+3\right) \lambda  x^2+\left(13 \lambda ^3+8\right) \lambda ^2 x-2\right)\nonumber\\
&&+1024 \lambda ^4 q^6 \left(109 \lambda ^5+3 x^5+69 \lambda  x^4-1032 \lambda ^2 x^3+154 \lambda ^3 x^2+523 \lambda ^4 x\right)\nonumber\\
&&-12288 \lambda ^5 q^4 \left(111 \lambda ^4+7 x^4-306 \lambda  x^3+364 \lambda ^2 x^2+477 \lambda ^3 x\right)\nonumber\\
&&+81920 \lambda ^6 q^2 \left(65 \lambda ^3-56 x^3+246 \lambda  x^2+267 \lambda ^2 x\right)-6881280 \lambda ^7 (\lambda +2 x)^2\Bigg]\nonumber\\
&&+\frac{1}{q^9}\lambda ^2 \left(2 q^6+27 q^4-600 q^2+1680\right) \log \left(4 \lambda +q^2 x+q \sqrt{\mc R_n}\right),
 \eea
 \bea
 \mc F^{\theta (2)}_f(r)&=&\frac{\lambda ^2 }{q^9}\Biggl[\frac{q \sqrt{\mc R_f} }{r}\Biggl(\frac{\left(q^2-12\right) q^2}{r^2}+\frac{264-62 q^2}{r}+\frac{8r^4 q^2 \left(q^6-q^4 (4 r+21)+q^2 (22 r+96)-32 (r+4)\right)}{\left(q^2-4\right) \mc R_f^2}\nonumber\\
 &&+\frac{2 \left(q^{10}+q^8 (1-2 r)-2 q^6 (39 r+238)+8 q^4 (119 r+568)-64 q^2 (57 r+244)+4608 (r+4)\right)}{\left(q^2-4\right)^2 \left(q^2+r (r+4)\right)}\Bigg)\nonumber\\
 &&-\left(2 q^6+27 q^4-600 q^2+1680\right) \log\frac{q^2 r+q \sqrt{\mc R_f}+2 r^2}{r^2}\Biggl].
 \eea
 Then those $p$-independent terms are obtained as
 \bea
 I^{\theta (2)}&=&\mc F^{\theta (2)}_f(r_f)-\mc F^{\theta (2)}_n(x_n)
 -\frac{\lambda ^2(\log\ep-2\log2q^2) }{q^9}\left(2 q^6+27 q^4-600 q^2+1680\right)\nonumber\\
 &&+\frac{q^3}{64 \lambda ^4}+\frac{1}{\lambda  q^3}
 +\frac{-4 q^{10}-21 q^8+2576 q^6-24288 q^4+83712 q^2-99584}{q^9 \left(q^2-4\right)^2}\lambda^2.
 \eea
 It is straightforward to push the calculation to the next order. So in the following we just list the results of $I^{(3)}$ and omit the details. At the NNNLO of the $\ep$ expansion, we have
 \bea\label{rtheta3}
 I^{\theta (3)}&=&\mc F^{\theta (3)}_f(r_f)-\mc F^{\theta (3)}_n(x_n)\nonumber\\
 &&+\frac{-591360 \lambda ^3+q^{10}+24 \left(4 \lambda ^3-3\right) q^8-80 \left(\lambda ^3-3\right) q^6-43680 \lambda ^3 q^4+322560 \lambda ^3 q^2}{4 q^{13}}(\log\ep-2\log2q^2)\nonumber\\
 &&+\frac{1}{384 \lambda ^6 q^{13} \left(q^2-4\right)^3}\Bigg(13974372352 \lambda ^9+q^{24}-12 q^{22}+9093513216 \lambda ^9 q^4-17877958656 \lambda ^9 q^2\nonumber\\
 &&+24 \left(\lambda ^3+2\right) q^{20}-48 \lambda ^3 \left(11 \lambda ^3-36\right) q^{16}-3072 \lambda ^6 \left(3665 \lambda ^3+908\right) q^{10}+36864 \lambda ^6 \left(7595 \lambda ^3+193\right) q^8\nonumber\\
 &&-16 \left(6 \lambda ^6-21 \lambda ^3-4\right) q^{18}-16384 \lambda ^6 \left(139750 \lambda ^3+417\right) q^6
 +192 \lambda ^3 \left(128 \lambda ^6-169 \lambda ^3-20\right) q^{14}\nonumber\\
 &&-512 \lambda ^3 \left(955 \lambda ^6-960 \lambda ^3-6\right) q^{12}\Biggl),
 \eea
 where

 \bea
 \mc F^{\theta (3)}_n(x)&=&\frac{1}{384 q^{13}}\Biggl[\frac{1}{\lambda ^6 \left(q^2-4\right)^3 \mc R_n^{5/2}}\Biggl(
 x^5 q^{29}+2 x^3 \left(-6 x^2+10 \lambda  x+5 \lambda ^2\right) q^{27}+\dots\Biggl)\nonumber\\
&&-96 \left(-591360 \lambda ^3+q^{10}+24 \left(4 \lambda ^3-3\right) q^8-80 \left(\lambda ^3-3\right) q^6-43680 \lambda ^3 q^4+322560 \lambda ^3 q^2\right) \nonumber\\
&&\times\log \left(4 \lambda +q \sqrt{\mc R_n}+q^2 x\right)\Biggl],
 \eea
 and
 \bea
 \mc F^{\theta (3)}_f(r)&=&\frac{1}{12 q^{13}}\Bigg[\frac{q \sqrt{\mc R_f}}{r}\Biggl(-\frac{32 \lambda ^3 q^4 \left(3 q^2-20\right)}{r^3}+\dots\Biggl)\nonumber\\
 &&+3 \left(-591360 \lambda ^3+q^{10}+24 \left(4 \lambda ^3-3\right) q^8-80 \left(\lambda ^3-3\right) q^6-43680 \lambda ^3 q^4+322560 \lambda ^3 q^2\right)\nonumber\\
&& \times\log \frac{q^2 r+q \sqrt{\mc R_f+2 r^2}}{r^2}\Biggl].
 \eea
 We find that the $\mc F$ functions at this order is too lengthy  to be listed even in the appendix, so we just put them in the ancillary
{\em Mathematica} files.

With the above expressions in hand, we are able to define a sum function as
\be\label{sumf}
S^{(i)}_{\omega}=\sum_{k=0}^{i}I^{\omega(k)}\epsilon^{k}
\ee
where, $i\in \mathbb{N}$ and $\omega\in\{\theta,\phi, t\}$. For example, we have
\be
S^{(3)}_{\theta}=\sum_{k=0}^{3}I^{\theta(k)}\epsilon^{k}=I^{\theta (0)}+I^{\theta (1)}\ep+I^{\theta (2)}\ep^2+I^{\theta (3)}\ep^3,
\ee
which could be viewed as an analytical approximation for the radial integrals $I^\theta$ in Eq. (\ref{raidalinteral_rtheta}).

In addition, in order to examine the accuracy of the approximate functions $S^{(i)}_{\omega}$, we would like to introduce the corresponding relative error functions as
\be\label{relf}
\delta^{(i)}_{\omega}=\frac{|S^{(i)}_\omega-I^{\omega}|}{I^{\omega}},
\ee
where $I^\omega$ denotes the exact value of the radial integrals of the $r-\omega$ motion.

\begin{figure}[h]
\begin{center}
\includegraphics[width=140mm,angle=0]{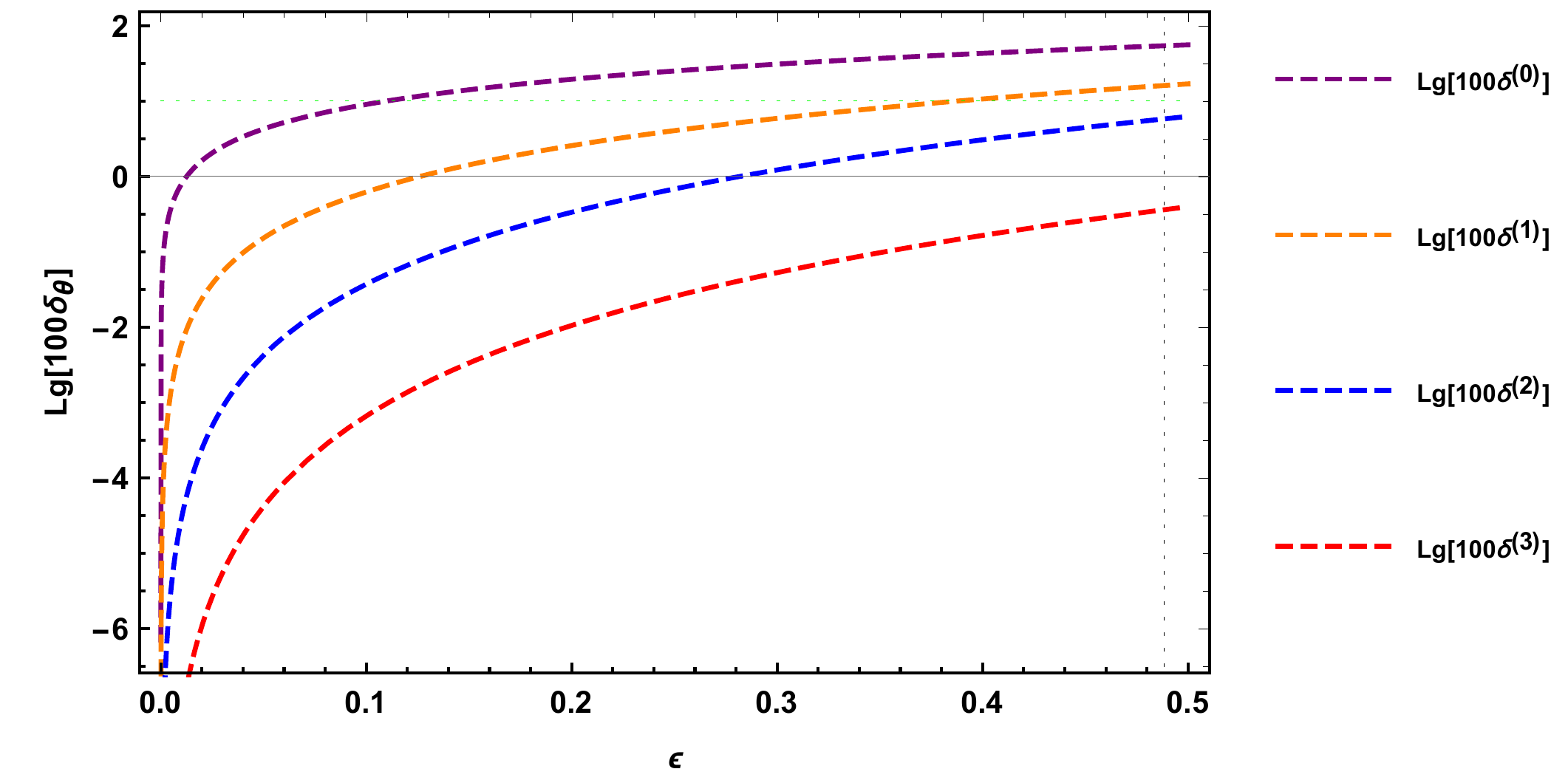}
\end{center}
\vspace{-5mm}
 \caption {The relative error function $\delta^{(i)}_\theta$ with respect to $\epsilon$ for $q=\sqrt3$ and $\lambda=1/2$. The green and black dotted horizon lines correspond to $\delta_\theta=10\%$ and $\delta_\theta=1\% $, respectively. The black dotted vertical line is $\epsilon=0.488$ which corresponds to $a=0.94$. For simplicity, we drop the subscript $\theta$ in the labels $\delta^{(i)}$ of the lines.}\label{rtheta}
\end{figure}

On the other hand, to choose a suitable value of $r_n$, we note that the ISCO radius can be expanded in $\ep$, which in terms of the coordinates (\ref{BHcoor}) is given by
 \be
 r_{ISCO}=2^{1/3}  \epsilon+\frac{7 }{ 2^{1/3}4} \epsilon ^2+\frac{15 }{32} \epsilon ^3+\mc O(\epsilon ^4).
 \ee
 One can see that the leading order of $r_{ISCO}$ is $\mc O(\ep)$, so if the emitter is moving on the ISCO then the light ray can extend to the far region.  Without loss of generality, here and in the subsequent sections we take $r_n=2^{1/3}\ep$, $r_f=100$ for the examination of the analytically approximate expressions for the radial integrals and we give the numerical results of $\delta_\theta^{(i)}$ in Fig. \ref{rtheta} at $q=\sqrt3$ and $\lambda=1/2$.

From this figure, we have to include the contributions up to the NNNLO term, i.e. $I^{\theta(3)}$ at least if we want to study the non-extremal Kerr black hole with $a=0.94$ ($\epsilon=0.488$) and retain the error within $1\%$. If we can tolerate an error of no more than 10 percent, pushing the results to the NNLO is enough. On the other hand, if only the leading term participates in the contribution, to retain the error within $1\%$, $\ep$ has to be smaller than 0.013, which means $a$ has to be larger than 0.999999. Even if we make the error be within less demanding standards, for example, the final results can have been put out by as much as $10\%$, $\ep$ has to be smaller than $0.12$.  Hence, in this sense, when $q$ is much bigger than $\ep$, the higher corrections could help the theoretical results closer to the observations of real black holes, such as M87*.

 \begin{figure}[h]
\begin{center}
\includegraphics[width=140mm,angle=0]{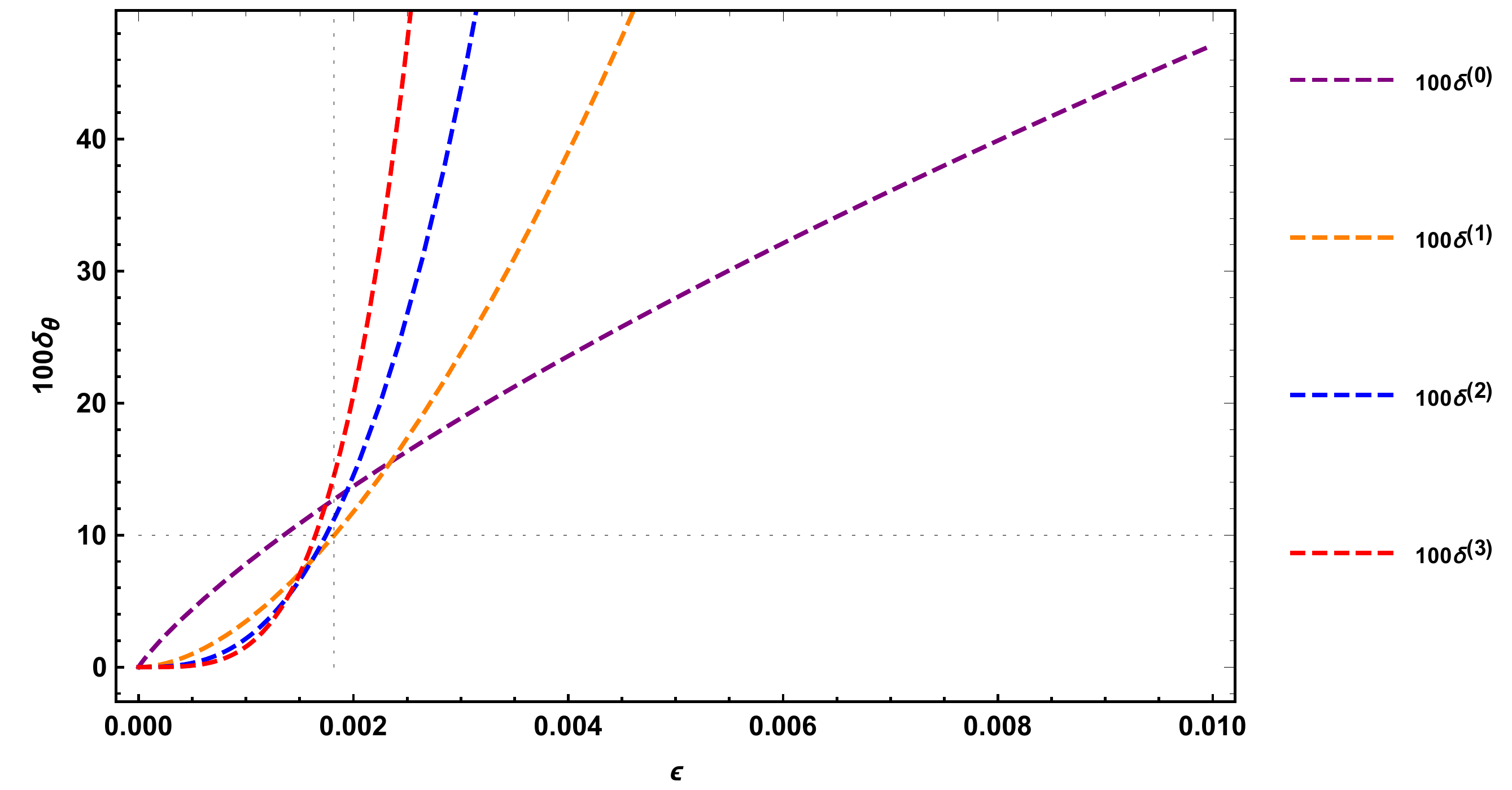}
\end{center}
\vspace{-5mm}
 \caption {The relative error function $\delta^{(i)}_\theta$ with respect to $\epsilon$ for $q=\lambda=1/2$. The black dotted horizon line is $\delta_\theta=10\%$. The black dotted vertical line is $\epsilon=0.0018$ which corresponds to $a=0.999999997$. For simplicity, we drop the subscript $\theta$ of the labels $\delta^{(i)}$ of the lines.}\label{rtheta1}
\end{figure}

On the other hand, we also draw the pictures in Fig. \ref{rtheta1} for $q=\lambda=1/2$. From this graph, we can find  $\ep\le0.0018$ if we allow the error to be within $10\%$, and the higher order terms could have advantages. However, if $\ep$ goes a little bigger, the higher order terms turn out to make things worse rather than better. This can be explained as follows. As we mentioned in the previous section, in the near-horizon region $r\simeq r_n\simeq \ep$ the validity of the expansion in $\ep$ requires that the parameter in the radial function $R(r)$ is much larger than $\ep$. In this case, the shifted Carter constant $q$ cannot take a small value if $\ep$ is not very small. On the other hand, the behavior in the far region $r\gg\ep$ is much better, in this case the validity of the high spin expansion is independent of the value of $\ep$.

Given all this, we conclude that the higher order terms can not only make those approximations more accurate, but also make it possible to calculate the radial integrals of null geodesics in the $r-\theta$ motion for $a\simeq 0.9$ when $q$ is large. But for a small $q$, $\ep$ has to be very small and it is impossible to extend to the case that the Kerr black hole is not near extreme, no matter whether the higher orders are considered or not.

Nevertheless, the story is not complete, we still need to check the radial integrals of the $r-\phi$ and $r-t$ motion.

\section{The radial integral for the $r-\phi$ motion}\label{section4}

In this section, we pay our attention to the radial integral
\be\label{rphiraidalinteral}
I^\phi=\int_{r_n}^{r_f}\frac{\Phi}{(r^2-\ep^3)\sqrt{R}}dr,
\ee
for the $r-\phi$ motion. As before, this integral can be separated into
\be
I^{\phi}=I_n^{\phi}+I_f^{\phi},
\ee
with
\be
I_n^{\phi}=\int^{\ep^p}_{r_n} \frac{\Phi}{(r^2-\ep^3)\sqrt{R}}dr,\quad
I_f^{\phi}=\int^{r_f}_{\ep^p} \frac{\Phi}{(r^2-\ep^3)\sqrt{R}}dr.
\ee
Then the integrals in the two regions can be analytically performed  respectively in the high spin expansion and then via the matched asymptotic expansion method the solution in the whole region can be obtained.

\begin{figure}[h]
\begin{center}
\includegraphics[width=140mm,angle=0]{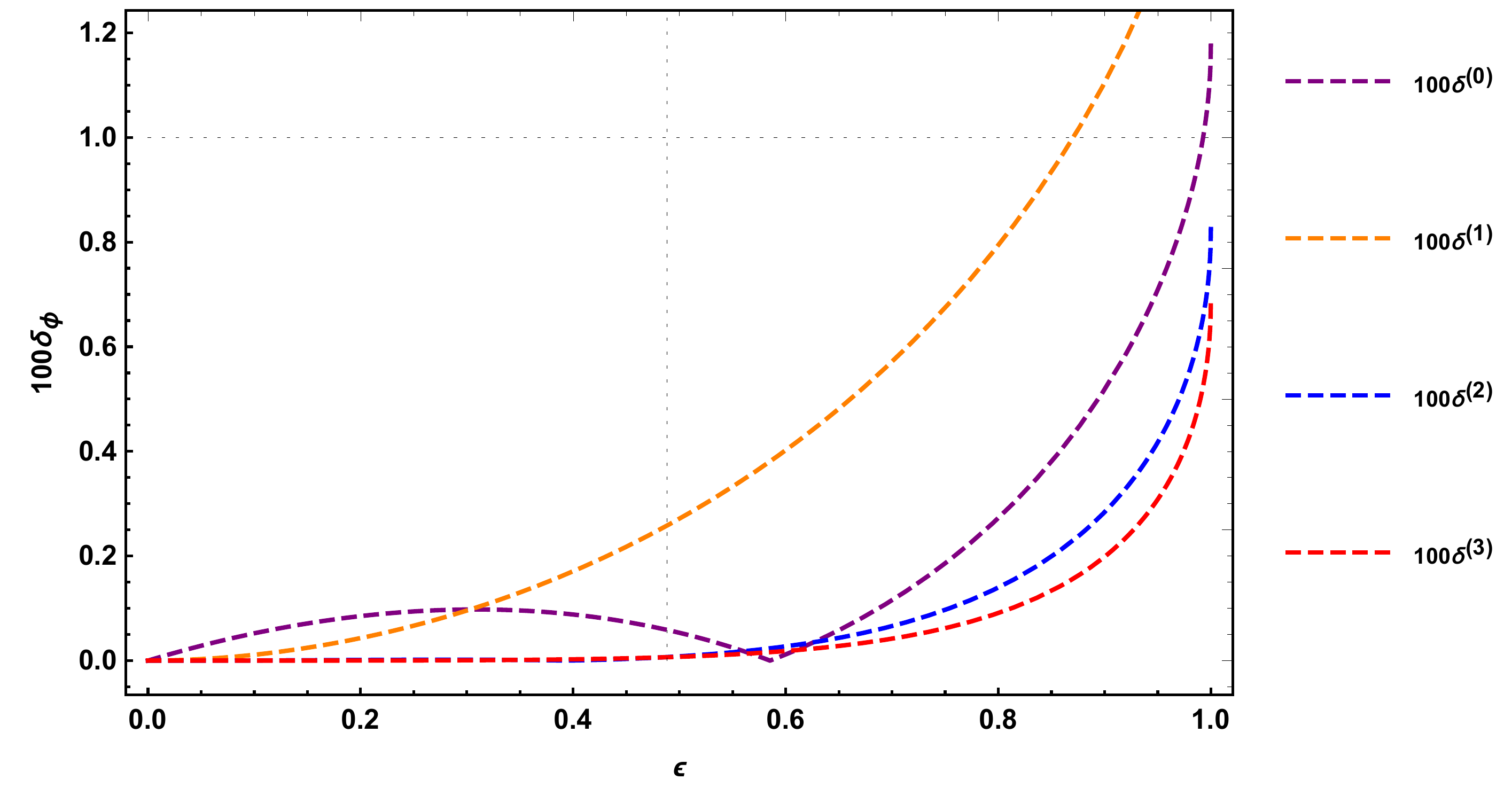}
\end{center}
\vspace{-5mm}
 \caption {The relative error function $\delta^{(i)}_\phi$ with respect to $\epsilon$  for $q=\sqrt{3}$ and $\lambda=1/2$. Also, the black dotted horizon line is $\delta_\theta=1\% $. The black dotted vertical line is $\epsilon=0.488$ which corresponds to $a=0.94$. For simplicity, we drop the subscript $\phi$ of the labels $\delta^{(i)}$ of the lines.}\label{rphi}
\end{figure}

\begin{figure}[h]
\begin{center}
\includegraphics[width=170mm,angle=0]{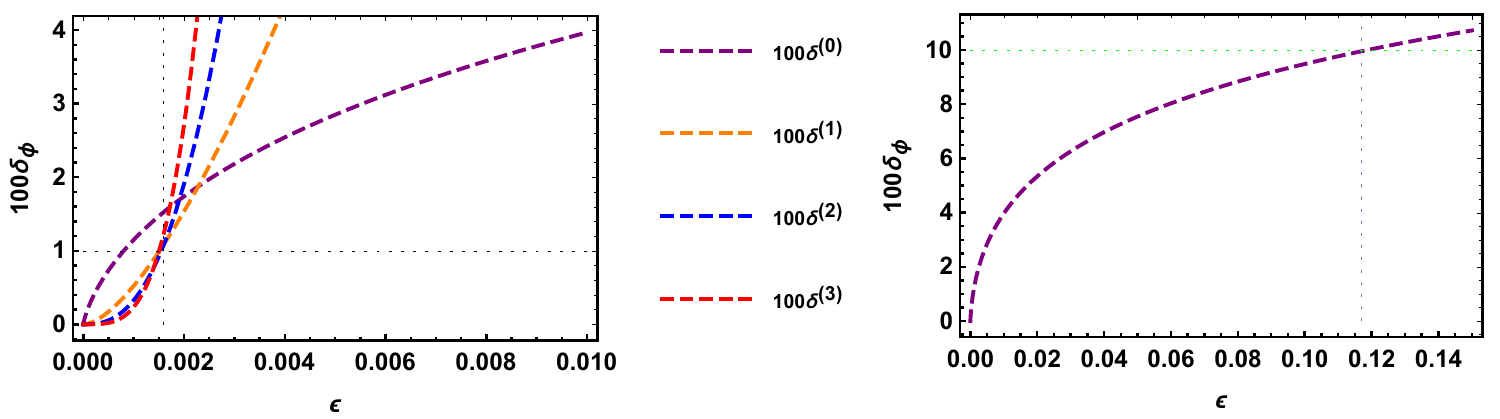}
\end{center}
\vspace{-9mm}
 \caption {The relative error function $\delta^{(i)}_\phi$ with respect to $\epsilon$  for $q=\lambda=1/2$. In the left graph, the green and black dotted horizon lines are $\delta_\theta=10\%$ and $\delta_\theta=1\% $, respectively. The grey dotted vertical line is $\epsilon=0.488$ which corresponds to $a=0.94$. For simplicity, we drop the subscript $\phi$ in the labels $\delta^{(i)}$ of the lines. In the right graph, we plot $\delta^{(0)}_\phi$ in the range $\ep\in\left(0,0.15\right)$ to show the maximum value of $\ep$ to keep the error tolerance to be less than $10\%$.}\label{rphi1}
\end{figure}

At LO, we have
\be
I_n^{\phi(0)}=\mc F^{\phi(0)}_n(\ep^{p-1})-\mc F^{\phi(0)}_n(x_n),
\ee
and
\be
I_f^{\phi(0)}=\mc F^{\phi(0)}_f(r_f)-\mc F^{\phi(0)}_f(\ep^p),
\ee
where
\be
\mc F^{\phi(0)}_n(x)=\frac{3 \log \left(4 \lambda +q \sqrt{\mc R_n}+q^2 x\right)}{q}-2\log\frac{2 \lambda +\sqrt{q^2 x^2+4 \lambda  (\lambda +2 x)}+2 x}{x},
\ee
\be
\mc F^{\phi(0)}_f(r)=-\frac{3 }{q}\log\frac{q \sqrt{\mc R_f}+q^2r+2 r^2}{r^2}+2\log\frac{\sqrt{\mc R_f}+r^2+2r}{r}+\frac{\sqrt{\mc R_f}}{r}.
\ee
The matching in the overlap region (\ref{verlapregion}) with the  integrals at higher orders of $\ep$ cancels the $p$-dependent terms and gives
\be\label{LOIphi}
I^{\phi(0)}=\mc F^{\phi(0)}_f(r_f)-\mc F^{\phi(0)}_n(x_n)+
\frac{6 \log 2 q^2}{q}-\frac{3 \log \epsilon }{q}-q-4 \log (q+2).
\ee
At NLO, similarly we have
\be
I_n^{\phi(1)}=\mc F^{\phi(1)}_n(\ep^{p-1})-\mc F^{\phi(1)}_n(x_n),
\ee
and
\be
I_f^{\phi(1)}=\mc F^{\phi(1)}_f(r_f)-\mc F^{\phi(1)}_f(\ep^p).
\ee
Then the integral in the whole region at this order is given by
\bea
I^{\phi(1)}&=&\mc F^{\phi(1)}_f(r_f)-\mc F^{\phi(1)}_n(x_n)-\frac{4 \lambda  \left(q^2-6\right) \left(q^2-3\right) (\log\epsilon-2\log2q^2 )}{q^5}
\nonumber\\
&&+\frac{7680 \lambda ^3+3 q^8-4 \left(8 \lambda ^3+3\right) q^6+1024 \lambda ^3 q^4-5376 \lambda ^3 q^2}{8 \lambda ^2 q^5 \left(q^2-4\right)}.
\eea
From now on,  we move all the higher order $\mc F^{\phi(i)}_n(x)$ and $\mc F^{\phi(i)}_f(r)$ functions with $i=1,2,3$ to the appendix \ref{appendixA}.
In the same way, at NNLO, we find
\be
I_n^{\phi(2)}=\mc F^{\phi(2)}_n(\ep^{p-1})-\mc F^{\phi(2)}_n(x_n),
\ee
and
\be
I_f^{\phi(2)}=\mc F^{\phi(2)}_f(r_f)-\mc F^{\phi(2)}_f(\ep^p).
\ee
The radial integral at this order is given by
\bea
I^{\phi(2)}&=&\mc F^{\phi(2)}_f(r_f)-\mc F^{\phi(2)}_n(x_n)+\frac{3 \lambda ^2(\log \epsilon-2\log2q^2) }{q^9} \left(4 q^6-187 q^4+1080 q^2-1680\right)\nonumber\\
&&+\frac{1}{64 \lambda ^4 q^9 \left(q^2-4\right)^2}
\Bigg(-19120128 \lambda ^6+3 q^{16}-24 q^{14}-9029632 \lambda ^6 q^4+21430272 \lambda ^6 q^2\nonumber\\
&&-64 \lambda ^3 \left(\lambda ^3-27\right) q^{10}
-64 \lambda ^3 \left(1975 \lambda ^3+72\right) q^8+1024 \lambda ^3 \left(1673 \lambda ^3+3\right) q^6\nonumber\\
&&+16 \left(8 \lambda ^6-12 \lambda ^3+3\right) q^{12}\Bigg).
\eea
Again, at NNNLO, we obtain
\be
I_n^{\phi(3)}=\mc F^{\phi(3)}_n(\ep^{p-1})-\mc F^{\phi(3)}_n(x_n),
\ee
and
\be
I_f^{\phi(3)}=\mc F^{\phi(3)}_f(r_f)-\mc F^{\phi(3)}_f(\ep^p).
\ee
The integral is then given by
\bea\label{rphi3}
I^{\phi(3)}&=&\mc F^{\phi(3)}_f(r_f)-\mc F^{\phi(3)}_n(x_n)\nonumber\\
&&+\frac{\log\ep-2\log2q^2}{4 q^{13}}\Bigg(-1774080 \lambda ^3-2 q^{12}-381920 \lambda ^3 q^4+1451520 \lambda ^3 q^2\nonumber\\
&&+\left(99-32 \lambda ^3\right) +q^{10}+168 \left(2 \lambda ^3-3\right) q^8+720 \left(43 \lambda ^3+1\right) q^6\Bigg)\nonumber\\
&&+\frac{1}{384 \lambda ^6 q^{13} \left(q^2-4\right)^3}\Bigg(41923117056 \lambda ^9+3 q^{24}-64977371136 \lambda ^9 q^2\nonumber\\
&&+4 \left(2 \lambda ^3-9\right) q^{22}-24 \left(7 \lambda ^3-6\right) q^{20}-147456 \lambda ^6 \left(94346 \lambda ^3+139\right) q^6+41480355840 \lambda ^9 q^4\nonumber\\
&&-48 \left(14 \lambda ^6-23 \lambda ^3+4\right) q^{18}-1024 \lambda ^6 \left(232825 \lambda ^3+16092\right) q^{10}+12288 \lambda ^6 \left(208571 \lambda ^3+2361\right) q^8\nonumber\\
&&+192 \lambda ^3 \left(1616 \lambda ^6-3611 \lambda ^3-12\right) q^{14}-16 \lambda ^3 \left(576 \lambda ^6-2829 \lambda ^3+140\right) q^{16}\nonumber\\
&&+1536 \lambda ^3 \left(4489 \lambda ^6+3072 \lambda ^3+6\right) q^{12}\Bigg).
\eea
With the above expressions, we have the final result,
\be
S^{(i)}_{\phi}=\sum_{k=0}^{i}I^{\phi(k)}\epsilon^{k},
\ee
and
\be
\delta^{(i)}_{\phi}=\frac{|S^{(i)}_\phi-I^{\phi}|}{I^{\phi}}.
\ee
We plot the relative error functions in Fig. \ref{rphi} and \ref{rphi1} for $q=\sqrt3$ and $\lambda=1/2$ and $q=\lambda=1/2$, respectively.

From these figures, we find that it is better to include the contributions of higher order terms only if $\ep$ is small enough, no matter $q$ is large or small. When the deviation $\ep$ gets to a certain size and then increases, considering a higher order will raise the relative error, that is, the leading term gives the best approximate values of the original integrals. In addition, we find the analytical expressions are not sensitive to $\ep$ for a large $q$, even $\ep$ increases to $0.9$, the approximate results are still controlled within $1\%$ error. However, turning to a small $q$ and looking at the the best analytical expression, that is, the leading term, the relative error can quickly exceed $10$ percent corresponding to $\ep=0.118$ and $a=0.9992$ which is far from the real rotating black hole in our universe.

\section{The radial integral for the $r-t$ motion}\label{section5}
In this section, let us finish our calculation with the radial integral
\be
I^t=\int_{r_n}^{r_f}\frac{T(r)}{(r^2-\ep^3)\sqrt{R(r)}}dr,\label{raidalinteral_rt}
\ee
for the $r-t$ motion, which can be separated into
\be
I^{t}=I_n^{t}+I_f^{t},
\ee
with
\be
I_n^{t}=\int^{\ep^p}_{r_n} \frac{T}{(r^2-\ep^3)\sqrt{R}}dr,\quad
I_f^{t}=\int^{r_f}_{\ep^p} \frac{T}{(r^2-\ep^3)\sqrt{R}}dr.
\ee
At LO, we have
\be
I_n^{t(0)}=\mc F^{t(0)}_n(\ep^{p-1})-\mc F^{t(0)}_n(x_n),
\ee
and
\be
I_f^{t(0)}=\mc F^{t(0)}_f(r_f)-\mc F^{t(0)}_f(\ep^p),
\ee
where
\be
\mc F^{t(0)}_n(x)=-\frac{\sqrt{\mc R_n}}{\lambda  x }\frac{1}{\epsilon },
\ee
\bea
\mc F^{t(0)}_f(r)&=&\frac{\sqrt{\mc R_f} }{r}\left(1-\frac{4}{q^2 r}\right)-\frac{\left(7 q^2-8\right) }{q^3}\log \left(\frac{q^2 r+q \sqrt{\mc R_f}+2 r^2}{r^2}\right)\nonumber\\
&&+2 \log \left(\frac{r^2+2 r+\sqrt{\mc R_f}}{r}\right).
\eea
Then the integral is given by
\be\label{rt0}
I^{t(0)}=\mc F^{t(0)}_f(r_f)-\mc F^{t(0)}_n(x_n)-\frac{q}{\lambda  \epsilon }.
\ee
At NLO, we have
\be
I_n^{t(1)}=\mc F^{t(1)}_n(\ep^{p-1})-\mc F^{t(1)}_n(x_n),
\ee
and
\be
I_f^{t(1)}=\mc F^{t(1)}_f(r_f)-\mc F^{t(1)}_f(\ep^p).
\ee
Then the integral is given by
\bea
I^{t(1)}&=&\mc F^{t(1)}_f(r_f)-\mc F^{t(1)}_n(x_n)\nonumber\\
&&+\frac{1}{\ep}\Bigg(-\frac{q^3}{12 \lambda ^3}+\frac{24}{q^3}+\frac{\left(8-7 q^2\right) (\log \epsilon-2\log2q^2 )}{q^3}
-q-\frac{4}{q}-4 \log (q+2)\Bigg).
\eea
Similar to the previous section, we have moved all the  higher order $\mc F^{t(i)}_n(x)$ and $\mc F^{t(i)}_f(r)$ functions with $i=1,2,3$ to the appendix \ref{appendixA}.
At NNLO, we have
\be
I_n^{t(2)}=\mc F^{t(2)}_n(\ep^{p-1})-\mc F^{t(2)}_n(x_n),
\ee
and
\be
I_f^{t(2)}=\mc F^{t(2)}_f(r_f)-\mc F^{t(2)}_f(\ep^p).
\ee
Then the integral is given by
\bea
I^{t(2)}&=&\mc F^{t(2)}_f(r_f)-\mc F^{t(2)}_n(x_n)\nonumber\\
&&+\frac{6 \lambda  (\log \epsilon-2\log2q^2 )}{\ep q^7} \left(9 q^4-52 q^2+80\right)\nonumber\\
&&-\frac{1}{80\ep \lambda ^5 q^7 \left(q^2-4\right)}\Bigg(558080 \lambda ^6+q^{14}-4 q^{12}-30 \lambda ^3 q^{10}+40 \left(4 \lambda ^6+\lambda ^3\right) q^8\nonumber\\
&&-80 \lambda ^3 \left(163 \lambda ^3-4\right) q^6+136000 \lambda ^6 q^4-480000 \lambda ^6 q^2\Bigg).
\eea
At NNNLO, we have
\be
I_n^{t(3)}=\mc F^{t(3)}_n(\ep^{p-1})-\mc F^{t(3)}_n(x_n),
\ee
and
\be
I_f^{t(3)}=\mc F^{t(3)}_f(r_f)-\mc F^{t(3)}_f(\ep^p),
\ee
Then the integral is given by
\bea\label{rt3}
I^{t(3)}&=&\mc F^{t(3)}_f(r_f)-\mc F^{t(3)}_n(x_n)\nonumber\\
&&-\frac{3 \lambda ^2(\log\ep-2\log2q^2) }{\ep q^{11}}\left(4 q^8+183 q^6-2640 q^4+10640 q^2-13440\right)\nonumber\\
&&-\frac{1}{448\ep \lambda ^7 q^{11} \left(q^2-4\right)^2}\Bigg(-1102839808 \lambda ^9+q^{22}-8 q^{20}-224 \lambda ^3 q^{16}+1399078912 \lambda ^9 q^2\nonumber\\
&&+\left(7 \lambda ^3+16\right) q^{18}+448 \lambda ^6 \left(1263 \lambda ^3+136\right) q^{10}-21504 \lambda ^6 \left(841 \lambda ^3+5\right) q^8-692142080 \lambda ^9 q^4\nonumber\\
&&-112 \lambda ^3 \left(8 \lambda ^6-13 \lambda ^3-13\right) q^{14}+448 \lambda ^3 \left(57 \lambda ^6-35 \lambda ^3-6\right) q^{12}+14336 \lambda ^6 \left(11485 \lambda ^3+6\right) q^6\Bigg).\nonumber\\
\eea

 \begin{figure}[h]
\begin{center}
\includegraphics[width=140mm,angle=0]{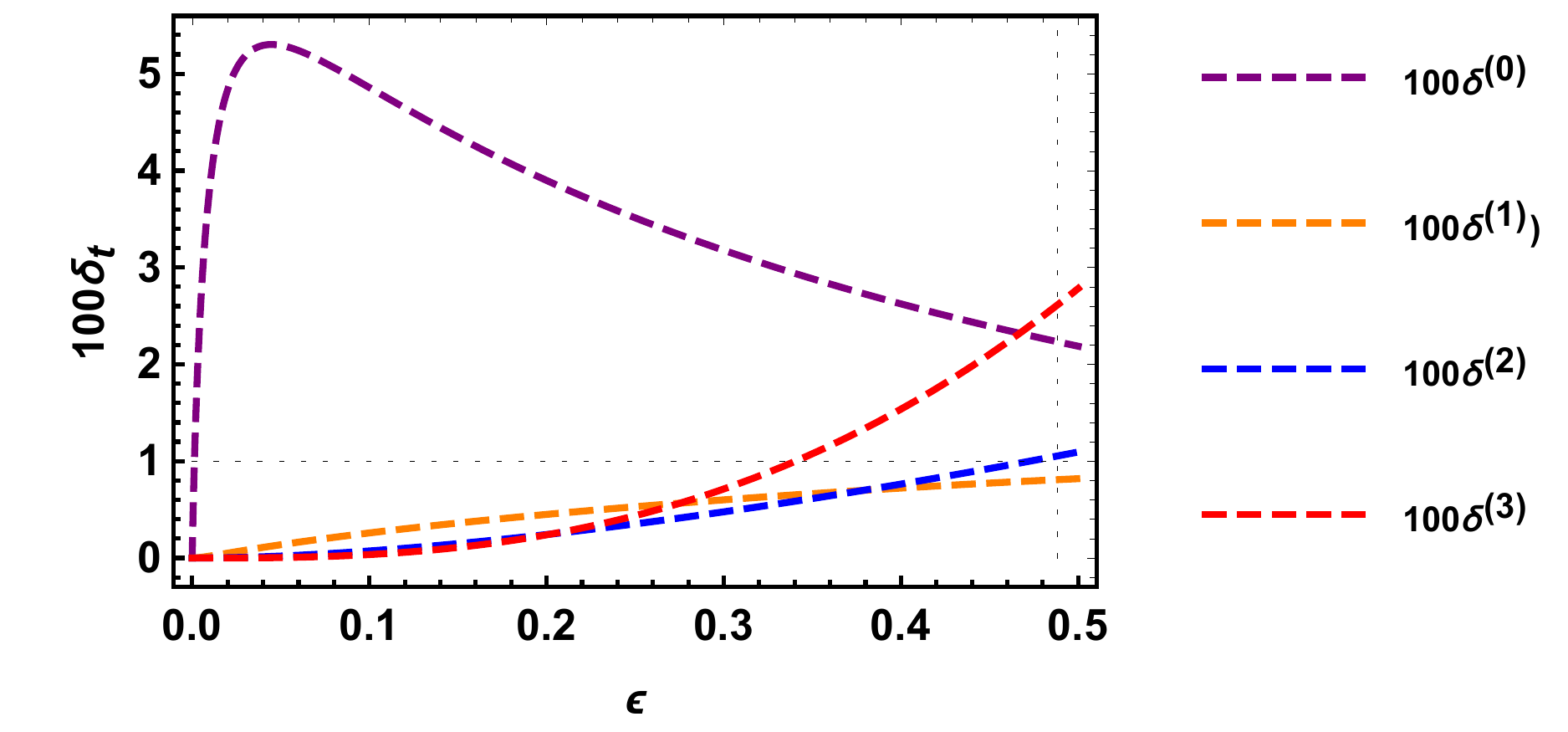}
\end{center}
\vspace{-5mm}
 \caption {The relative error function $\delta^{(i)}_t$ with respect to $\epsilon$  for $q=3/2$ and $\lambda=1/2$. The black dotted horizon line is $\delta_\theta=1\% $, respectively. The grey dotted vertical line is $\epsilon=0.488$ which corresponds to $a=0.94$. For simplicity, we drop the subscript $\phi$ of the labels $\delta^{(i)}$ of the lines.}\label{rt}
\end{figure}

 \begin{figure}[h]
\begin{center}
\includegraphics[width=140mm,angle=0]{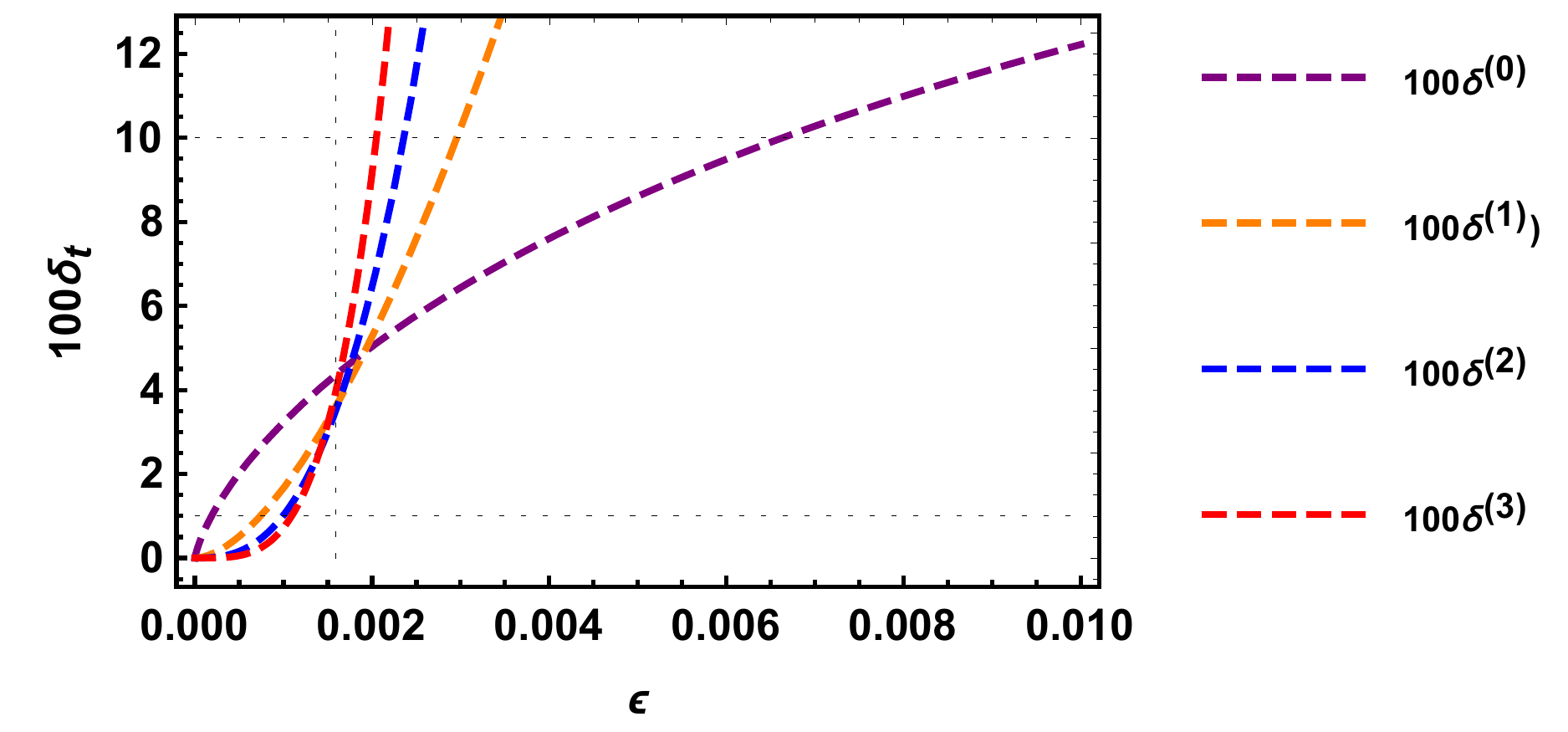}
\end{center}
\vspace{-5mm}
 \caption {The relative error function $\delta^{(i)}_t$ with respect to $\epsilon$  for $q=\lambda=1/2$.The black dotted horizon lines are $\delta_\theta=10\%$ and $\delta_\theta=1\% $, respectively. The black dot-dashed vertical line is $\epsilon=0.0016$ which corresponds to $a=0.999999998$. For simplicity, we drop the subscript $\phi$ of the labels $\delta^{(i)}$ of the lines.}\label{rt1}
\end{figure}

As we did before, we present two examples for the large and small parameters $q$ in Fig. \ref{rt} and \ref{rt1}. For the small $q$ case, we choose  $q=\lambda=1/2$ again, while unlike the other motions we choose  $q=3/2$ and $\lambda=1/2$ to take a glance at the difference that changing $q$ makes when $q$ is relatively large.

From Fig. \ref{rt}, we find that the expressions including the leading and next-to-leading terms are the best approximations of the  original integrals when $\ep$ is as large as $1/2$. The relative error is always less than $1\%$ in the range $\ep\in(0,1/2)$. Compared with the results of the other motions, we conclude that when beyond a certain order, in contrast to our expect, the  incorporation of more higher order terms would decrease the accuracy of the approximate expressions. Turning to the small $q$, the approximate expressions are also shown to apply only when the $\ep$ is very small, which means the high spin expansion and MAE methods do not work well if one wants to find the analytical expressions instead of the numerical calculations approximatively.

\section{Summary}\label{summary}
Due to the emergence of an enhanced symmetry in the near-horizon geometry of the (near-) extreme Kerr black hole, the radial integrals of the null geodesic equations can be analytically obtained \cite{Porfyriadis:2016gwb}. Following the method proposed in \cite{Porfyriadis:2016gwb} and \cite{Gralla:2017ufe}, we extended the computation of the radial integrals of null geodesics to higher orders in $\ep$, where $\ep$ is a very small parameter characterizing the spin of the Kerr black hole deviating from extremality $a=M$.

Considering the null geodesics that begin from the near-horizon region $r_n\ll 1$ and end at the far region $r_f\gg\ep$. We first expanded the radial integrals in terms of $\ep$ for both regions, in particular, as pointed out in \cite{Gralla:2017ufe}, we also made the change of variable $x=r/\ep$ in the near-horizon region to ensure the validity of the expansion. By employing the method of MAE, we obtained the analytical expressions of the radial integrals up to NNNLO in $\ep$. The results of the first four orders for $r-\theta$ motion can be found in Eqs. (\ref{rtheta0})-(\ref{rtheta3}). We showed the expressions for $r-\phi$ motion in Eqs. (\ref{LOIphi})-(\ref{rphi3}). And for $r-t$ motion, the results are given in Eqs. (\ref{rt0})-(\ref{rt3}). As expected, our results of the leading term are the same with those in \cite{Porfyriadis:2016gwb,Gralla:2017ufe}.

By comparing these analytical expressions with the numerical evaluations of the exact integrals, we investigated the effectiveness of the high spin expansion method. To achieve this, we defined the sum function in Eq. (\ref{sumf}) and the relative error function in Eq. (\ref{relf}). We found that the validity of the high spin expansion requires that the shifted Carter constant $q\gg\ep$. For a large $q$, the analytical expressions behave very well, thus the results  may be applied to the study of the electromagnetic signature in the spacetime of M87* black hole whose spin is supposed to be $a\simeq0.9 M$. However, for a small $q$, to retain a satisfactory precision our results show that $\ep$ has to be very small. In this case, the high spin expansion method can only apply to (near-) extreme black holes, excluding black holes like M87*.  New methods have to be developed to solve this problem. Nevertheless, when it comes the situation with a large $q$, the results from the high spin expansion method may be a good choice due to their analyticity and accuracy.

In \cite{Gralla:2019ceu}, the authors obtained the complete solutions of the null geodesic equations in the Kerr spacetime, which are valid for any value of the black hole spin and any value of the photon angular momentum. Their results are expressed in terms of elliptic functions. More recently, in \cite{Gates:2020sdh}, the authors found at the leading order of the high spin expansion, the radial integral $I^\theta$ can be recovered as an asymptotic expansion of the exact results of \cite{Gralla:2019ceu}. It would be interesting to explore the possibility of recovering the results in this paper from the exact solutions obtained in \cite{Gralla:2019ceu} \footnote{We are thankful to anonymous referee for pointing out this point.}. We leave this to future work.

\section*{Acknowledgments}
We thank Peng-Xiang Hao, Jiang Long  and Haopeng Yan for useful discussions. The work is in part supported by NSFC Grant No. 11335012, No. 11325522 and No. 11735001.MG and PCL are also supported by NSFC Grant No. 11947210. MG is also funded by China Postdoctoral Science Foundation Grant No. 2019M660278 and 2020T130020. PCL is also funded by China
Postdoctoral Science Foundation Grant No. 2020M670010.

\appendix
\section{Details of $\mathcal{F}^\phi$ and $\mathcal{F}^t$ functions}\label{appendixA}
In this appendix, we show all the higher order $\mathcal{F}^{\phi(i)}_n(x)$ and $\mathcal{F}^{\phi(i)}_f(r)$ functions of the radial integrals  for the $r-\phi$ motion and  $\mathcal{F}^{t(i)}_n(x)$ and $\mathcal{F}^{t(i)}_f(r)$ functions of the radial integrals  for the $r-t$ motion:
\bea
\mc F^{\phi(1)}_n(x)&=&\frac{1}{8 \lambda ^2 q^5 \left(q^2-4\right) x^2\sqrt{\mc R_n} }\Biggl(3 q^9 x^3+2304 \lambda ^3 q x^2 (\lambda +2 x)+192 \lambda ^2 q^3 x^2 \left(-8 \lambda ^2+x^2-17 \lambda  x\right)\nonumber\\
&&+4 q^7 \left(-4 \lambda ^3+8 \lambda ^2 x^4-3 x^3+3 \lambda  x^2-4 \lambda ^2 x\right)+16 \lambda  q^5 (4 \lambda ^2-11 \lambda  x^4+34 \lambda ^2 x^3\nonumber\\
&&+\left(16 \lambda ^3-3\right) x^2+4 \lambda  x)+
32 \lambda ^3 \left(q^6-13 q^4+54 q^2-72\right) \sqrt{\mc R_n} x^2 \log \left(4 \lambda +q^2 x+q \sqrt{\mc R_n}\right)\Biggl),\nonumber\\
\eea
\bea
\mc F^{\phi(1)}_f(r)&=&-\frac{4 \lambda  \left(q^2-6\right) \left(q^2-3\right) }{q^5}\log \frac{q^2 r+q \sqrt{\mc R_f}+2 r^2}{r^2}\nonumber\\
&&+\frac{4 \lambda  \left(q^2-3\right) \left(q^4 (r-1)-2 q^2 \left(2 r^2+9 r-2\right)+12 r (r+4)\right)}{q^4 \left(q^2-4\right)  \sqrt{\mc R_f}},
\eea
\bea
\mc F^{\phi(2)}_n(x)&=&\frac{1}{64 \lambda ^4 q^8 \left(q^2-4\right)^2 x^4 \mc R_n^{3/2}}\Biggl(3 q^{18} x^7-20643840 \lambda ^7 x^4 (\lambda +2 x)^2-6 q^{16} x^5 \left(-3 \lambda ^2+4 x^2-6 \lambda  x\right)\nonumber\\
&&-12288 \lambda ^5 q^4 x^4 \left(693 \lambda ^4+21 x^4-1238 \lambda  x^3+2452 \lambda ^2 x^2+2911 \lambda ^3 x\right)
-245760 \lambda ^6 q^2 x^4 (-89 \lambda ^3\nonumber\\
&&+56 x^3-342 \lambda  x^2-363 \lambda ^2 x)+
1024 \lambda ^4 q^6 x^4 (1391 \lambda ^5+9 x^5+279 \lambda  x^4-6152 \lambda ^2 x^3\nonumber\\
&&+3918 \lambda ^3 x^2+6153 \lambda ^4 x)
++8 q^{14} x^2 (-12 \lambda ^5+4 \lambda ^4 x^7+\left(6-24 \lambda ^3\right) x^5\nonumber\\
&&-12 \lambda  \left(4 \lambda ^3+3\right) x^4-\lambda ^2 \left(16 \lambda ^3+9\right) x^3+9 \lambda ^3 x^2-12 \lambda ^4 x)
-16 \lambda  q^{12} (16 \lambda ^6+35 \lambda ^3 x^9\nonumber\\
&&+96 \lambda ^4 x^8-12 \lambda ^2 \left(2 \lambda ^3+9\right) x^7+4 \left(8 \lambda ^6-37 \lambda ^3-9\right) x^6+2 \lambda  \left(52 \lambda ^3+9\right) x^5\nonumber\\
&&+2 \lambda ^2 \left(64 \lambda ^3+21\right) x^4+8 \lambda ^3 \left(4 \lambda ^3-7\right) x^3-24 \lambda ^4 x^2+48 \lambda ^5 x)
-256 \lambda ^3 q^8 (16 \lambda ^4+37 \lambda  x^9\nonumber\\
&&+465 \lambda ^2 x^8-2 \left(2279 \lambda ^3+6\right) x^7-4 \lambda  \left(\lambda ^3-11\right) x^6+3 \lambda ^2 \left(523 \lambda ^3+56\right) x^5\nonumber\\
&&+2 \left(152 \lambda ^6+64 \lambda ^3+3\right) x^4+8 \lambda  \left(4 \lambda ^3-1\right) x^3+24 \lambda ^2 x^2+48 \lambda ^3 x)\nonumber\\
&&
+64 \lambda ^2 q^{10} (32 \lambda ^5+55 \lambda ^2 x^9+345 \lambda ^3 x^8-12 \lambda  \left(109 \lambda ^3+6\right) x^7-8 \left(52 \lambda ^5+\lambda ^2\right) x^6\nonumber\\
&&+\left(-32 \lambda ^7+256 \lambda ^4+30 \lambda \right) x^4+8 \lambda ^2 \left(8 \lambda ^3-5\right) x^3+24 \lambda ^3 x^2+96 \lambda ^4 x)\nonumber\\
&&+\left(-48 \lambda ^6+304 \lambda ^3+18\right) x^5\Biggl)\nonumber\\
&&-\frac{3 \lambda ^2 \log \left(4 \lambda +q \sqrt{\mc R_n}+q^2 x\right)}{q^9} \left(4 q^6-187 q^4+1080 q^2-1680\right),\nonumber\\
\eea
\bea
\mc F^{\phi(2)}_f(r)&=&\frac{\lambda ^2 }{q^9}
\Biggl[\frac{\sqrt{q\mc R_f}}{r}\Biggl(\frac{\left(11 q^2-36\right) q^2}{r^2}+\frac{40 q^4-394 q^2+792}{r}\nonumber\\
&&-\frac{8 r^4 q^2 \left(q^2-3\right) \left(q^6-q^4 (4 r+21)+q^2 (22 r+96)-32 (r+4)\right)}{\left(q^2-4\right) \mc R_f^2}\\
&&-\frac{1}{\left(q^2-4\right)^2 \mc R_f}\Bigg(2 r^2(q^{12}-2 q^{10} (r+3)-q^8 (56 r+379)+2 q^6 (517 r+2626)\nonumber\\
&&-8 q^4 (753 r+3400)+64 q^2 (235 r+988)-13824 (r+4))\Bigg)\Biggl)\nonumber\\
&&+3 \left(4 q^6-187 q^4+1080 q^2-1680\right) \log \left(\frac{q^2 r+q \sqrt{\mc R_f}+2 r^2}{r^2}\right)\Biggl],
\eea
\bea
\mc F^{\phi(3)}_n(x)&=&\frac{ \log \left(4 \lambda +q \sqrt{\mc R_n}+q^2 x\right)}{4 q^{13}}\Bigg(1774080 \lambda ^3+2 q^{12}+381920 \lambda ^3 q^4-1451520 \lambda ^3 q^2\nonumber\\
&&+\left(32 \lambda ^3-99\right) q^{10}-168 \left(2 \lambda ^3-3\right) q^8-720 \left(43 \lambda ^3+1\right) q^6\Bigg)\nonumber\\
&&+\frac{1}{384 \lambda ^6 q^{13} \left(q^2-4\right)^3 \mc R_n^{5/2} x^6}\Bigg(3 q^{29} x^{11}+174399160320 \lambda ^{11} q x^6 (\lambda +2 x)^3+\dots\Bigg),\nonumber\\
\eea
\bea
\mc F^{\phi(3)}_f(r)&=&\frac{\sqrt{\mc R_f}}{12 q^{12}r}
\Bigg(\frac{32 \lambda ^3 q^4 \left(q^4-21 q^2+60\right)}{r^3}+\dots\Bigg)\nonumber\\
&&-\frac{1}{4 q^{13}}\log \left(\frac{q^2 r+q \sqrt{\mc R_f}+2 r^2}{r^2}\right)\Bigg(1774080 \lambda ^3+2 q^{12}+\left(32 \lambda ^3-99\right) q^{10}+\dots\Bigg),\nonumber\\
\eea
\bea
\mc F^{t(1)}_n(x)&=&\frac{1}{12 \lambda ^3 q^3 \sqrt{\mc R_n} x^3 \epsilon }\Bigg(24 \lambda ^3 q^3 \sqrt{\mc R_n} x^3 \log (x)+12 \lambda ^3 \left(7 q^2-8\right) \sqrt{\mc R_n} x^3 \log \left(4 \lambda +q^2 x+q \sqrt{\mc R_n}\right)\nonumber\\
&&-q\left(q^6 x^4+2 \lambda  q^4 x^2 (\lambda +2 x)+8 \lambda ^2 q^2 \left(2 \lambda ^2+6 \lambda  x^4-x^2+2 \lambda  x\right)-96 \lambda ^3 x^3 (\lambda +2 x)\right)\nonumber\\
&&+24 \lambda ^3 q^2 \sqrt{\mc R_n} x^3 \log \left(\sqrt{\mc R_n}+2 (\lambda +x)\right)\Bigg),
\eea
\bea
\mc F^{t(1)}_f(r)&=&-\frac{2 \lambda }{q^7 \left(q^2-4\right) r \sqrt{\mc R_f}}\Bigg(-2 q^9 r^2+2 q^9 r+q^9+8 q^7 r^3+51 q^7 r^2-32 q^7 r-8 q^7\nonumber\\
&&-102 q^5 r^3-468 q^5 r^2+136 q^5 r+16 q^5+392 q^3 r^3+1648 q^3 r^2-160 q^3 r\nonumber\\
&&-3 \left(9 q^6-88 q^4+288 q^2-320\right) r \sqrt{\mc R_f} \log \left(\frac{q^2 r+q \sqrt{\mc R_f}+2 r^2}{r^2}\right)-480 q r^3-1920 q r^2\Bigg),
\nonumber\\
\eea
\bea
\mc F^{t(2)}_n(x)&=&-\frac{1}{80 \lambda ^5 q^6 \left(q^2-4\right) \mc R_n^{3/2} x^5 \epsilon }
\Bigg(q^{16} x^8+2 q^{14} x^6 \left(3 \lambda ^2-2 x^2+6 \lambda  x\right)+614400 \lambda ^7 x^5 (\lambda +2 x)^2\nonumber\\
&&+10240 \lambda ^6 q^2 x^5 \left(-49 \lambda ^3+40 x^3-186 \lambda  x^2-201 \lambda ^2 x\right)-6 \lambda  q^{12} x^4 \left(-\lambda ^3+5 \lambda ^2 x^4+8 x^3-4 \lambda ^2 x\right)\nonumber\\
&&+7680 \lambda ^5 q^4 x^5 \left(17 \lambda ^4+x^4-46 \lambda  x^3+56 \lambda ^2 x^2+73 \lambda ^3 x\right)
-8 \lambda ^2 q^{10} x^2 (-12 \lambda ^4+40 \lambda ^3 x^7\nonumber\\
&&+5 \lambda  \left(4 \lambda ^3-1\right) x^6+45 \lambda ^2 x^5+\left(12-25 \lambda ^3\right) x^4-4 \lambda  \left(5 \lambda ^3-4\right) x^3+9 \lambda ^2 x^2-12 \lambda ^3 x)\nonumber\\
&&-128q^6\lambda ^4(8 \lambda ^4+55 \lambda  x^9-810 \lambda ^2 x^8+4 \lambda  \left(5 \lambda ^3-1\right) x^3+12 \lambda ^2 x^2+24 \lambda ^3 x\nonumber\\
&&+5 \left(7 \lambda ^3-6\right) x^7+5 \lambda  \left(71 \lambda ^3-16\right) x^6+5 \lambda ^2 \left(16 \lambda ^3+3\right) x^5+\left(60 \lambda ^3+3\right) x^4)\nonumber\\
&&+32 \lambda ^3 q^8 (8 \lambda ^5+80 \lambda ^2 x^9-5 \left(59 \lambda ^3-2\right) x^8-5 \lambda  \left(32 \lambda ^3-3\right) x^7-5 \lambda ^2 \left(4 \lambda ^3+21\right) x^6\nonumber\\
&&+\left(4-5 \lambda ^3\right) x^5+\left(60 \lambda ^4+9 \lambda \right) x^4+4 \lambda ^2 \left(5 \lambda ^3-4\right) x^3+24 \lambda ^4 x)\Bigg)\nonumber\\
&&-\frac{6 \lambda   \log \left(4 \lambda +q^2 x+q \sqrt{\mc R_n}\right)}{q^7 \epsilon }\left(9 q^4-52 q^2+80\right),
\eea
\bea
\mc F^{t(2)}_f(r)&=&-\frac{3 \lambda ^2 \left(4 q^8+183 q^6-2640 q^4+10640 q^2-13440\right) }{q^{11}}\log  \left(\frac{q^2 r+q \sqrt{\mc R_f}+2 r^2}{r^2}\right)\nonumber\\
&&+\frac{\lambda ^2 \sqrt{\mc R_f}}{q^{10} r}\Bigg[\frac{8 \left(q^2-4\right) q^4}{r^3}+\frac{\left(23 q^4-268 q^2+544\right) q^2}{r^2}
+\frac{2 \left(8 q^6-445 q^4+2564 q^2-3936\right)}{r}\nonumber\\
&&-\frac{8 q^2 \left(q^{10}-4 q^8 (r+7)+q^6 (54 r+275)-18 q^4 (15 r+68)+32 q^2 (19 r+80)-512 (r+4)\right)}{\left(q^2-4\right) \left(q^2+r (r+4)\right)^2}\nonumber\\
&&-\frac{2}{\left(q^2-4\right)^2 \left(q^2+r (r+4)\right)}\Bigg(q^{14}-2 q^{12} (r+9)+98304 (r+4)-q^{10} (20 r+223)\nonumber\\
&&+2 q^8 (639 r+3686)-8 q^6 (1651 r+8052)+64 q^4 (917 r+4080)-512 q^2 (239 r+1000)\Bigg)\Bigg],\nonumber\\
\eea
\bea
\mc F^{t(3)}_n(x)&=&\frac{3 \lambda ^2 \log \left(4 \lambda +q \sqrt{\mc R_n}+q^2 x\right)}{\ep q^{11}}
\left(4 q^8+183 q^6-2640 q^4+10640 q^2-13440\right)\nonumber\\
&&+\frac{1}{448\ep \lambda ^7 q^{10} \left(q^2-4\right)^2 x^7 \mc R_n^{5/2}}\Bigg(4624220160 \lambda ^{11} x^7 (\lambda +2 x)^3-q^{26} x^{12}+\dots\Bigg),\nonumber\\
\eea
\bea
\mc F^{t(3)}_f(r)&=&-\frac{1}{12 q^{15}}\Bigg[\frac{q \sqrt{\mc R_f}}{r}\Bigg(-\frac{24 \lambda ^3 q^6 \left(q^4-24 q^2+80\right)}{r^4}+\dots\Biggl)\nonumber\\
&&+3\log \left(\frac{q^2 r+q \sqrt{\mc R_f}+2 r^2}{r^2}\right)\Bigg(-15375360 \lambda ^3+2 q^{14}-91 q^{12}+\dots\Bigg)\Bigg].
\eea
The explicit form of the $\mc F$ functions at the NNNLO can be found in the ancillary {\em Mathematica} files.

\end{document}